\documentstyle[12pt]{ioplppt}                

\begin{document}

\jl{5}

\title{Differential constraints for the Kaup -- Broer system as a reduction
of the 1D Toda lattice}

\author{A K Svinin\ftnote{1}{E-mail address : svinin@icc.ru}}

\address{Institute of System Dynamics and Control Theory,
Siberian Branch of Russian Academy of Sciences,
P.O. Box 1233, 664033 Irkutsk, Russia}

\begin{abstract}
It is shown that some special reduction of infinite 1-D Toda lattice
gives differential constraints compatible with the Kaup -- Broer
system. A family of the travelling wave solutions of the Kaup --
Broer system and its higher version is constructed.
\end{abstract}

\maketitle

\section{Introduction}

The goal of this paper is to show that differential constraints
compatible with the Kaup -- Broer system \cite{kaup}, \cite{broer}
\begin{equation}
\eqalign{
\displaystyle
\frac{\partial S}{\partial t_2} = - S_{xx} + 2SS_x + 2R_x, \\
\frac{\partial R}{\partial t_2} = R_{xx} + 2(RS)_x
}
\label{BK}
\end{equation}
can be derived as a special reduction of the 1D Toda lattice and to select
some class of travelling wave solutions
of it and its higher counterpart.

It is well known that the Kaup -- Broer system is applied in hydrodynamics.
It appears as a model equation for nonlinear water waves.
In turn, the Kaup -- Broer hierarchy appears to be related with
one-Hermitian matrix model
and can be extracted from Toda lattice hierarchy where the first flow
parameter is treated as the space coordinate \cite{bonora}.

As is known, the hierarchy of the Kaup -- Broer system is
intimately related to the
Kadomtsev -- Petviashvili hierarchy and can be interpreted as a
special reduction of it whose evolution equations are coded by the
Lax equation
\begin{equation}
\frac{\partial {\cal L}}{\partial t_p} = [{\cal L}_{+}^p, {\cal L}]
\label{Lax}
\end{equation}
on the monic pseudodifferential operator
${\cal L} = \partial + \sum_{i=1}^{\infty}U_i(x)\partial^{-i}$.
The corresponding Lax operator is constrained by the condition
\[
{\cal L} = \partial +
\sum_{i=1}^{\infty}U_i[S, R]\partial^{-i} = \partial + R(\partial - S)^{-1},\;\;
\partial = \frac{\partial}{\partial x}.
\]

The relationship with the Toda lattice hierarchy provides evolution equations
of the Kaup -- Broer hierarchy by the property of the
existence of discrete symmetries
generated by similarity transformation
\begin{equation}
\overline{{\cal L}} = (\partial - S^0){\cal L}(\partial - S^0)^{-1}
\label{similarity}
\end{equation}
where $S^0 = S + \partial\ln R$. More explicitly, we have
\begin{equation}
\overline{S} = S + \frac{R_x}{R},\;\;
\overline{R} = R + S_x + \frac{R_{xx}}{R} - \frac{R_x^2}{R^2}.
\label{explicitly}
\end{equation}

One of the effective methods of searching for exact solutions is to select
them by means of differential constraints \cite{sidorov}, \cite{andreev}.
In this article we find the differential constraints for the Kaup -- Broer system
and its higher counterpart. To this
aim, we use a reduction of the
Toda lattice which is among many found recently in
\cite{svinin}.
These constraints select a broad class of traveling wave solutions.

This paper is organized as follows. In section 2, we recall definition
of the 1D Toda lattice and its relationship with the Kaup -- Broer hierarchy.
In section 3, we exhibit a denumerable class of reductions of the infinite
1D Toda lattice \cite{svinin}. Next, we discuss the case $n=1$. We show
that in this case the reduced system serves as differential constraints
for the Kaup -- Broer system and its higher counterpart. Also, we show
that these differential constraints isolate the solutions of travelling
wave type.

\section{Preliminaries}

We give here
background information on the 1D Toda lattice and its relationship with
the Kaup -- Broer hierarchy. Also we recall the notion of differential
constraints compatible with a given evolutionary system of partial
differential equations.

The Toda lattice, as is known, can be represented as the consistency condition
of linear auxiliary equations
\begin{equation}
\eqalign{
L(\psi_i) = \psi_{i+1} + a_0(i)\psi_i + a_1(i)\psi_{i-1} = z\psi_i, \\
\psi_{ix} = A(\psi_i) = \psi_{i+1} + a_0(i)\psi_i,\;\;
i\in{\bf Z}.
}
\label{auxiliary}
\end{equation}
The corresponding Lax equation $\partial(L) = [A, L]$ is
equivalent to equations
of the one-dimensional Toda lattice
\begin{equation}
\eqalign{
\displaystyle
a_{0x}(i) = a_1(i+1) - a_1(i), \\
a_{1x}(i) = a_1(i)(a_0(i) - a_0(i-1)).
}
\label{TL1}
\end{equation}
Introducing variables $u_i$ by
\begin{equation}
a_0(i) = - u_{ix},\;\;
a_1(i) = e^{u_{i-1} - u_i}
\label{u_i}
\end{equation}
one can rewrite the Toda lattice (\ref{TL1}) in usual form:
\begin{equation}
u_{ixx} =  e^{u_{i-1} - u_i}  - e^{u_i - u_{i+1}}.
\label{TL2}
\end{equation}

From (\ref{auxiliary}) one can easy derive that each wavefunction
$\psi_i$ satisfies the linear equation
\[
{\cal L}_i\psi_i = z\psi_i
\]
where
\[
{\cal L}_i = \partial + a_1(i)(\partial - a_0(i-1))^{-1}.
\]
It is easy to show that, by virtue of (\ref{TL1})
the operators ${\cal L}_i$ are related by the invertible gauge transformation
\begin{equation}
{\cal L}_{i+1} = (\partial - a_0(i)){\cal L}_i(\partial - a_0(i))^{-1}.
\label{gauge}
\end{equation}
Fixing any value $i = i_0 \in{\bf Z}$ one identifies
\begin{equation}
a_1(i_0+1) = R,\;\;
a_0(i_0) = S.
\label{RS}
\end{equation}
From The equations of Toda lattice (\ref{TL1})
one can easily extract the symmetry transformation
(\ref{explicitly}) generated by  the shift $i\rightarrow i+1$.

Let us also briefly discuss the notion of differential constraints
compatible with a given system of partial differential equations.
Let $E$ be a partial differential system with respect to functions
$u^1,..., u^m$ of two variables, say $t\in{\bf R}^1$ and $x\in{\bf R}^1$.
The notation
$[E]$ stands for union of $E$ and its differential consequences
with respect to $x$. In what follows we restrict ourselves by
consideraton only evolutonary equations
\begin{equation}
u^i_t = F^i[u^1,..., u^m],
\label{evolutionary}
\end{equation}
where $F^i$ are some (analytic) differential
functions of $u^1,..., u^m$. Let the system (\ref{evolutionary}) be
supplemented by differential constraints $H$
\begin{equation}
h_j[u^1,..., u^m] = 0,\;\;
j = 1,..., p,\;\;
p\leq m.
\label{constraints}
\end{equation}

One says that the differential constraints (\ref{constraints}) define
an invariant manifold of the system (\ref{evolutionary}) if
\begin{equation}
D_t(h_j)|_{[E]\cap[H]} = 0,\;\;
j = 1,..., p,
\label{determine}
\end{equation}
where $D_t$ denotes total derivative with respect to $t$.
Equations (\ref{determine}), whose solutions are some collections
of differential functions $\{h_1,..., h_p\}$, are reffered to as
determining ones.

It is quite difficult to use determining equations (\ref{determine}) with
$h_j$ in general form. However, the situation is considerably simplified
if (\ref{constraints}) can be resolved with respect to higher-order
derivatives as
\[
(u^j)_x^{(N_j)}=S^j(t, x, u^1,..., u^m, u_x^1, u_x^2,...).
\]
In this case a simple practical recipe to solve the determining equations
(\ref{determine})  consists of successively replacing
$(u^j)_x^{(N_j)}\rightarrow S^j$.

It is common of knowledge that the problem of finding all differential
constraints compatible with a given equation (system of equations) can be more
complicated than solving these equations. In practice, it is better to restrict
oneself to finding differential constraints in some fixed classes. Regarding
to the development of regular methods for constructing differential constraints,
see \cite{andreev}, \cite{kaptsov}.

\section{Reductons of the 1D Toda lattice, differential constraints
and travelling wave solutions of the Kaup -- Broer system}

Recently, we have proposed an infinite class of reduction of the Toda lattice
\cite{svinin}.
They are specified by constraints
\begin{equation}
\eqalign{
- a_0(i) - ... - a_0(i+n-1) = a_1(i)a_1(i+1)...a_1(i+n), \\
i\in{\bf Z},\;\;
n\in{\bf N}
}
\label{constraint1}
\end{equation}
or
\begin{equation}
u_{ix} + ... + u_{i+n-1,x} = e^{u_{i-1} - u_{i+n}}.
\label{constraint2}
\end{equation}

For any fixed $i=i_0\in{\bf Z}$, introduce a finite number of functions
$\{q_1,..., q_{n+1}\}$ identifying
\begin{equation}
q_1 = u_{i_0},\; q_2 = u_{i_0+1},..., q_{n+1} = u_{i_0+n}.
\label{q}
\end{equation}
From (\ref{TL2}) and (\ref{constraint2}) we derive the finite-dimensional
system
\begin{equation}
\begin{array}{l}
q_{1xx} = (q_{1x} + ... + q_{nx})e^{q_{n+1}-q_1} - e^{q_1-q_2},\\[0.3cm]
q_{kxx} = e^{q_{k-1}-q_k} - e^{q_k-q_{k+1}},\;\; k = 2,..., n,\\[0.3cm]
q_{n+1,xx} = e^{q_{n}-q_{n+1}} - (q_{2x} + ... + q_{n+1,x})e^{q_{n+1}-q_1}
\end{array}
\label{finite}
\end{equation}
togegher with discrete symmetry transformation generated by the shift
$i\rightarrow i+1$
\begin{equation}
\overline{q}_1 = q_2,..., \overline{q}_{n} = q_{n+1},\; \overline{q}_{n+1} = q_1 -
\ln\left[q_{2x} + ... + q_{n+1,x}\right].
\label{AB}
\end{equation}

In what follows, we restrict our attention to the case $n=1$. We have
\begin{equation}
q_{1xx} = q_{1x}e^{q_2-q_1} - e^{q_1-q_2},\;\;
q_{2xx} = e^{q_1-q_2} - q_{2x}e^{q_2-q_1}.
\label{system}
\end{equation}
Observe that the system (\ref{system}) can be cast into canonical Hamiltonian
setting. Generalized momenta are introduced as
\[
p_1 = - q_2^{\prime} - \frac{1}{2}e^{q_2 - q_1},\;\;
p_2 = - q_1^{\prime} - \frac{1}{2}e^{q_2 - q_1}.
\]
One can verify that equations (\ref{system}) are equivalent to
Hamiltonian system
\[
q_{ix} = \frac{\partial H}{\partial p_i},\;\;
p_{ix} = - \frac{\partial H}{\partial q_i},\;\;
i = 1, 2,
\]
where
\begin{equation}
H = - \left(p_1 + \frac{1}{2}e^{q_2 - q_1}\right)\left(p_2 +
\frac{1}{2}e^{q_2 - q_1}\right) + e^{q_1 - q_2}.
\label{H}
\end{equation}
The first two integrals of the (\ref{system}) are $H$ and $P = p_1 + p_2$.
It is simple exersice to check that $H$ and $P$ are in involution with respect
to standard Poisson bracket. So we can conclude that the equations
(\ref{system}) establish Hamiltonian system integrable in the sense of
Liouville theorem \cite{liouville}. It is natural to suppose that all systems
(\ref{finite}) are Liouville-integrable.

To proceed, we need to express variables $R$ and $S$ via $q_1$ and $q_2$.
Taking into account (\ref{u_i}), (\ref{RS}) and (\ref{q}) one obtains
\begin{equation}
\eqalign{
S = a_0(i_0) = - u_{i_0,x} = - q_{1x}, \\
\displaystyle
R = a_1(i_0+1) = e^{u_{i_0} - u_{i_0+1}} = e^{q_1-q_2}.
}
\label{express}
\end{equation}
From Toda lattice equations (\ref{TL1}), by virtue of (\ref{constraint1}),
we obtain the following differential equations:
\[
S_x = a_{0x}(i_0) = a_1(i_0+1) - a_1(i_0) = a_1(i_0+1) +
\frac{a_0(i_0)}{a_1(i_0+1)} = R + \frac{S}{R},
\]
\[
\overline{S}_x = a_{0x}(i_0+1) = a_1(i_0+2) - a_1(i_0+1)
\]
\[
= - \frac{a_0(i_0+1)}{a_1(i_0+1)}-a_1(i_0+1) = - \frac{\overline{S}}{R} - R
\]
where $\overline{S}$ is given by (\ref{explicitly}). The latter, as can be
checked, in more explicit form reads as a pair of ordinary differential
equations
\begin{equation}
\eqalign{
\displaystyle
S_x = R + \frac{S}{R}, \\
\displaystyle
R_{xx} = \frac{R_x^2}{R} - \frac{R_x}{R} - 2R^2 - 2S.
}
\label{system1}
\end{equation}
One can verify that differential substitution (\ref{express}) indeed maps
solutions of the system (\ref{system}) into solutions of (\ref{system1}).

By using determining equations (\ref{determine})
one can verify that equations (\ref{system1})
serve as differential constraints compatible with the Kaup -- Broer
system (\ref{BK}) and its higher version
\begin{equation}
\eqalign{
\displaystyle
\frac{\partial S}{\partial t_3} = S_{xxx} - 3SS_{xx} - 3S_x^2 +
6 (SR)_x + 3S^2S_x, \\
\displaystyle
\frac{\partial R}{\partial t_3} = R_{xxx} + 6RR_x + 3SR_{xx} + 3S_xR_x +
3(S^2R)_x.
}
\label{BK1}
\end{equation}
We conjecture that relations (\ref{system1}) play the role of differential
constraint for all members of the Kaup -- Broer hierarchy.
Following question arises: solutions of what kind are isolated by differential
constraints (\ref{system1})? The following proposition is helpful to
answer this question.

{\bf Proposition.} {\it By virtue of differential constraints
{\rm (\ref{system1})} following relations
hold:
\begin{equation}
\frac{\partial S}{\partial t_2} = PS_x,\;\;\;
\frac{\partial R}{\partial t_2} = PR_x,
\label{1}
\end{equation}
\begin{equation}
\frac{\partial S}{\partial t_3} = (E+P^2)S_x,\;\;\;
\frac{\partial R}{\partial t_3} = (E+P^2)R_x,
\label{2}
\end{equation}
where
\[ P = 2S + \frac{R_x}{R} - \frac{1}{R},
E = - S^2 + R - S\frac{R_x}{R}
\] are two first integrals of the system {\rm (\ref{system1})}.
} \\
This proposition is proved by straightforward calculation.

{\bf Remark 1.} The integrals $P$ and $E$ are $P = p_1 + p_2$ and $H$
(\ref{H}) expressed in terms of variables $S$ and $R$.

Taking into account the proposition above, it is natural to suppose that
there exists an infinite collection of polynomials $K_l(E, P)$ such
that by virtue (\ref{system1}), relations of the kind as in (\ref{1}) and (\ref{2})
are valid, i.e.
\begin{equation}
\frac{\partial S}{\partial t_l} = K_l(E, P)S_x,\;\;\;
\frac{\partial R}{\partial t_l} = K_l(E, P)R_x.
\label{3}
\end{equation}
Thus $K_1 = 1$, $K_2 = P$ and $K_3 = E+P^2$.

Next, we observe that, by virtue of (\ref{system1}), $E$ and $P$ also
do not depend on $t_l$. Take for example $E$. Taking into account
(\ref{3}) we have
\[
D_{t_l}(E) =
\frac{\partial E}{\partial R}R_xK_l +
\frac{\partial E}{\partial R_x}R_{xx}K_l +
\frac{\partial E}{\partial S}S_xK_l = D_x(E)K_l = 0,
\]
where $D_{t_l}$ and $D_x$ stands for total derivative with
respect to corresponding argument. The proposition above and this observation
prove that differential constraints (\ref{system1}) select simultaneous
solution of the systems (\ref{BK}) and (\ref{BK1}) in the form of
travelling wave
\begin{equation}
S = S(\xi),\;\;\;
R = R(\xi),
\label{wave}
\end{equation}
where $\xi = x + Pt_2 + (E+P^2)t_3 + \xi_0$, where $\xi_0$ is some constant
(it may depend on $t_4$, $t_5$,...).

Thus, to find a profile of travelling wave (\ref{wave}) which is
simultaneous
solution of (\ref{BK}) and (\ref{BK1}) we need to solve ordinary
differential equations
\begin{equation}
\eqalign{
\displaystyle
S^{\prime} = R + \frac{S}{R}, \\
\displaystyle
R^{\prime\prime} = \frac{R^{\prime 2}}{R} - \frac{R^{\prime}}{R} - 2R^2 - 2S.
}
\label{system2}
\end{equation}
with some initial conditions $(R_0 = R(0), R_1 = R^{\prime}(0), S_0 = S(0))$.

{\bf Remark 2.} In (\ref{wave}) $P$ and $E$ are understood as some values
$P = P_0$ and $E = E_0$ of first integrals corresponding to a particular
solution of (\ref{system2}).

To conclude this section, let us discuss Painlev\'{e} property for
system (\ref{system2}). Simple analysis shows that it passes
Painlev\'{e}
test. In addition, the system  (\ref{system2}) has a formal
``general" solution in the form of pole-like expansion
\[
S(\xi) = \frac{1}{\xi}\left(1 + c_1\xi + c_2\xi^2 - \frac{1}{2}\xi^3 -
(\frac{2}{5}c_1 + \frac{1}{5}c_2^2)\xi^4 + O(\xi^5)\right)
\]
\[
R(\xi) = \frac{1}{\xi^2}\left(-1 + c_2\xi^2 -
(\frac{1}{5}c_1 + \frac{3}{5}c_2^2)\xi^4 + O(\xi^5)\right)
\]
with two arbitrary constants $c_1$ and $c_2$. Fuchs indices are $-1$, $1$
and $2$.

Suppose now that solution
(\ref{wave}) does not depend on $t_2$. It is equivalent to assuming that
$P = 0$. In turn this requires that
\begin{equation}
S = - \frac{R^{\prime}}{2R} + \frac{1}{2R}.
\label{constr}
\end{equation}
It is easy to check that relation (\ref{constr}) properly defines
reduction of system (\ref{system2}) to the equation
\[
R^{\prime\prime} = \frac{R^{\prime 2}}{R} - \frac{1}{R} - 2R^2
\]
which is  particular case of Panlev\'{e} 12 equation \cite{ince}
\[
R^{\prime\prime} = \frac{R^{\prime 2}}{R} + \frac{\alpha}{R} + \beta
+ \gamma R^2 + \delta R^3
\]
with $\alpha = -1$, $\beta = 0$, $\gamma = -2$ and $\delta = 0$.

\section{Discussion}

In this paper we have derived differential constraints for
the Kaup -- Broer system via reducton of the infinite 1D Toda lattice.
It is shown that these differential constraints select a family of
travelling wave solutions.

Ablowitz -- Ramani -- Segur (ARS) conjecture states that any theoretical group
reduction of an integrable system of partial differential equations
will have the (generalized) Painlev\'e property \cite{ars}. We believe that
ARS conjecture could be extended on differential constraints.
The example of the system (\ref{system2}) exhibited in this paper supports
this conjecture.

\section*{Acknowledgments}

We are grateful to the referees for carefully reading the manuscript and for
their remarks which enabled us to improve the presentation of the paper.

The author wishes to thank the Editorial Board for the invitation to contribute
the paper to this issue. This research has been partially supported by
INTAS grant 2000-15.

\end{document}